\documentclass[journal]{IEEEtran}

\usepackage[utf8]{inputenc}
\usepackage{amsmath,amsfonts}
\usepackage{pifont}
\usepackage{algorithmic}
\usepackage{algorithm}
\usepackage{array}
\usepackage[caption=false,font=normalsize,labelfont=sf,textfont=sf]{subfig}
\usepackage{textcomp}
\usepackage{url}
\usepackage{verbatim}
\usepackage{graphicx}
\usepackage{cite}
\usepackage{diagbox}
\usepackage{graphicx}				
\usepackage{amssymb}
\usepackage{bm}
\usepackage{siunitx}
\usepackage{ifthen}
\usepackage[dvipsnames]{xcolor}   
\usepackage{hyperref}
\hypersetup{
	colorlinks=true, 
	linktoc=all,     
	linkcolor=blue,  
	citecolor=black,
}
\usepackage{cancel}
\usepackage{soul}

\newcommand{\myVec}[1]{\bm{#1}}
\newcommand{\vecx}{\myVec{x}}
\newcommand{\vece}{\hat{\myVec{e}}}
\newcommand{\veck}{{\myVec{k}}}
\newcommand{\vecU}{\myVec{U}}
\newcommand{\veckap}{{\boldsymbol \onewave}}
\newcommand{\veckcyl}{\veck_\mathrm{cyl}}
\newcommand{\vectwo}[2]{\begin{pmatrix}#1\\#2\end{pmatrix}}

\def\del{\partial}

\newcommand{\p}{\partial}
\newcommand{\pdiffOne}[2]{\frac{\del #1}{ \del #2}  }
\newcommand{\phih}{\hat{\phi}}
\newcommand{\vnabla}{\myVec{\nabla}}
\newcommand{\vecr}{\myVec{r}}
\newcommand{\DSV}{\tilde{c}}
\newcommand{\vecDSV}{\tilde{\myVec{c}}}
\newcommand{\myMean}[1]{\left\langle #1 \right\rangle}
\newcommand{\Angg}{ \theta}
\newcommand{\myvartheta}{ \Angg}
\newcommand{\Ang}{ \theta_0}
\newcommand{\AngU}{\Angg_U}
\newcommand{\FFT}{\mathrm{FFT}}
\newcommand{\DAng}{\Delta \Angg}
\newcommand{\AngDSV}{\vartheta_{\DSV}}
\newcommand{\ttheta}{\tilde{\theta}}
\newcommand{\kmin}{k_{\mathrm{min}}}
\newcommand{\kmax}{k_{\mathrm{max}}}
\newcommand{\omDR}{\omega_\text{DR}}
\newcommand{\half}{{\textstyle \frac12}}
\newcommand{\Uc}{\vecU_c}
\newcommand{\onewave}{\kappa}
\newcommand{\REFINDEX}{\text{ref}}
\newcommand{\kref}{k_\REFINDEX}
\newcommand{\omref}{\omega_\REFINDEX}
\newcommand{\Tref}{t_\REFINDEX}
\newcommand{\Lref}{l_\REFINDEX}
\newcommand{\CPref}{c_{p,\REFINDEX}}
\newcommand{\cp}{\text{cyl}}
\newcommand{\kcyl}{k_\cp}

\newcommand{\kCPx}{k_{\cp,x}}
\newcommand{\kCPy}{k_{\cp,y}}
\newcommand{\veckCP}{\veck_\cp}

\newcommand{\dcdk}{\delta c_{\delta k}}
\newcommand{\dcdw}{\delta c_{\delta \omega}}

\begin{document}
    \title{The effect of horizontal shear on extracting water currents from surface wave data}
 
	\author{Stefan Weichert, Benjamin K. Smeltzer, Simen \AA. Ellingsen 
	\thanks{SW and S{\AA}E were supported in part by the European Union (ERC, WaTurSheD, project 101045299) and S{\AA}E also by the Research Council of Norway (project 325114). Views and opinions expressed are, however, those of the authors only and do not necessarily reflect those of the European Union or the European Research Council. Neither the European Union nor the granting authority can be held responsible for them.}
	\thanks{S.~Weichert and S.~\AA.\ Ellingsen are with the Department of Energy and Process Engineering, Norwegian University of Science and Technology, 7491 Trondheim, Norway.} 
	\thanks{B.~K.\ Smeltzer is with SINTEF Ocean, Marinteknisk senter, 7052 Trondheim, Norway}}

	\maketitle
	
	\begin{abstract}
        The dispersive motion of surface waves is now routinely used to remotely measure the currents 
        close beneath
        the surface of oceans and other natural flows. The current manifests as wavelength-dependent Doppler shifts in the spatiotemporal wave spectrum, which is obtained by performing a Fourier transform of an observed wave field, a procedure which assumes that the current is horizontally uniform within the field of view. This assumption is frequently not satisfied. We here analyze the effects of the presence of horizontal shear in the velocity field, with special emphasis on biases: ``measured'' velocities which differ systematically in magnitude and/or direction compared to the average current within the field of view. We generate random synthetic wave data according to a spectrum prescribed at the domain center while varying the current and its gradient, and the mean direction $\theta_0$ and angular spread of the waves' propagation direction, from which we extract Doppler shift velocities (DSVs) using one of the most common numerical schemes. Assuming currents vary by up to 10\% of a characteristic wave phase velocity across the field of view, we find that strong biases can occur for highly directional wave spectra, strongly dependent on $\theta_0$ and the current direction relative to that of its magnitude gradient. We discuss how biases arise in the DSV extraction procedure and suggest practical steps to discover the presence of biases.
	\end{abstract}


\section{Introduction}

   Knowledge of near-surface currents in oceans, lakes and along coastlines is essential for a number of purposes, e.g.\ predicting the motion of water masses, oil spills, nutrients, microplastics and other pollutants. A highly attractive alternative to \emph{in situ} current measurements is to observe the surface wave spectrum from radar or visual data (e.g, \cite{Stewart74,Young85,Fernandez96,Gangeskar02,Ardhuin09,Hessner14,Lund15, stole-hentschel23
    } and \cite{Stresser17,Dugan03,Yurovskaya18,Lenain23}, respectively), and analyze the Doppler shifts induced by the current, compared to what would be the case in quiescent water (see \cite{Smeltzer21} for a review). Recent use of aircraft \cite{Lenain23} and drones \cite{Stresser17} demonstrate how the method allows for large water surfaces to be covered in short time, as opposed to single points or trajectories. 
    These examples also show a high spatial resolution in addition to the large coverage, e.g., \SI{1}{km} along-path resolution on flight paths $\sim \SI{100}{km}$ long \cite{Lenain23}, and $\SI{8}{m}$ resolution in a $\sim 190\times\SI{320}{m^2}$ area \cite{Stresser17}.

The method of current retrieval is based on performing a Fourier transform of the time-resolved surface shape inside an image window in both time and horizontal space before analyzing the spectrum on surfaces in wave vector-frequency space. This assumes, however, that the current is the same in the whole window, i.e., that it is horizontally uniform. Naturally, this is not always the case: in many locations where velocity fields are particularly desired the current can have sharp gradients, for instance ocean fronts \cite{Lenain23}, rivers \cite{Stresser17}, estuaries \cite{Lund18} and coastal tidal flow \cite{Cochin08}. 
Approaches which decompose the surface signal using localized basis functions (\emph{viz} wavelets) have been demonstrated \cite{Chernyshov22}, and data-driven methods show promise (e.g.\ \cite{Hackett21}), but Fourier transform-based analysis remains the standard, with the obvious advantage of a direct link to linear wave theory.

When horizontal current variations are present, in breach of the underlying assumptions of the method, one would expect the method to return the average value of the current within the image window; a value different from the average would constitute a measurement bias. 
In this article we study under which conditions the presence of horizontal current shear can lead to biases in the measured currents compared to the window average, and give estimates of the severity.

\section{Background}\label{sec:background}

Reconstruction of the vertical velocity profile is based on the wave vector-frequency spectrum obtained from surface elevation data resolved in spatial position $\vecx=(x,y)$ and time $t$, whence the linear-wave dispersion relation $\omDR(k_x,k_y)$ is extracted. The difference between the measured $\omDR$ and that predicted without current is the current-induced Doppler shift, from which the current can be reconstructed.

\subsection{Linear wave propagation without horizontal shear} \label{sec:waveTheoryNoshear}

This section is a synoptic recapitulation of the standard scenario; many more details may be found in, e.g., \cite{Smeltzer21}.
Remotely sensing the subsurface current using surface waves as a  probe makes use of the fact that the wave vector 
$\veck=(k_x,k_y)$ and frequency $\omega$ follow the dispersion relation
\begin{equation}\label{eq:dispRel}
    \omDR(\veck;\vecDSV) 
    = \omega_0(k)+\veck\cdot \vecDSV,
\end{equation}
where $\vecDSV$ is the Doppler shift velocity (DSV)
because of a background current $\vecU$ 
causing deviation of $\omDR$ from the quiescent water frequency $\omega_0=\sqrt{gk}$, where $k=|\veck|$ and $g$ is gravitational acceleration.
We consider only depth-uniform currents in this paper; for the simplest case where $\vecU$ is constant also horizontally, an ideal method should extract $\vecDSV=\vecU$ for all $\veck$. 
When the background current varies vertically 
but not horizontally (as is the underlying assumption behind the method), 
$\vecDSV$ is a function of $\veck$ and $\vecU(z)$; the measured values of $\vecDSV(\veck)$ can be inverted to extract $\vecU(z)$ (e.g.\ \cite{Smeltzer21}). 

We mention in passing that although $\vecU$ does not vary with depth in our examples, the \emph{biases} in the extracted DSV due to horizontal shear do vary with $\veck$, and hence a depth variation can be ``measured'' when none is present. 
        
The method described relies on using a 2D Fourier transform to obtain a spectrum in $\veck$ space. When the current is horizontally uniform within the window,
the spectral intensity is concentrated at, or near, the dispersion relation $\omega = \omega_\text{DR}$. 
If the current varies horizontally, however, the method's underlying assumption is violated: the local Doppler shift varies with position and there no longer exists a single value of $\vecDSV$.

\subsection{Linear waves on slowly varying currents} \label{sec:waveTheoryHshear}

Assume linear waves atop a current $\vecU(\vecx)$. We will generate synthetic wave data from a given input wave spectrum.
Assuming $\vecU$ varies slowly with $\vecx$ it can be taken to be locally uniform, i.e., the dispersion relation \eqref{eq:dispRel} may be used, with $\vecDSV=\vecU(\vecx)$.
The theory is well-established; see, e.g., \cite{Urs60,Pere}.

Each partial wave of the spectrum is
described by a cosine with argument (wave phase) $\phi(\vecx,t)
= \omega t - \phih(\vecx)$ where $\omega$ is constant for a steady current.
The local wave vector $\veck(\vecx)$ is then $\veck =-\vnabla \phih$.
Defining $\Omega(\vecx,\veck)=\omega_\text{DR}(\veck;\vecDSV(\vecx))$, the relation $\Omega=\omega$ is 
a  first order, nonlinear partial differential equation for $\phih(\vecx)$
which can be solved using the method of characteristics, yielding an ordinary differential equation for the variables $\vecr(t)$, $\phih(t)$, $\veck(t)$,
\begin{equation}\label{eq:rayeq}
    \pdiffOne{}{t}
    \begin{pmatrix}\vecr\\ \phih \\\veck\end{pmatrix}=
    \left.
    \begin{pmatrix}
    \vnabla_{\veck}\Omega \\
    \myVec{k}\cdot \vnabla_{\veck}\Omega\\ -\vnabla
    \Omega
    \end{pmatrix}
    \right|_{\myVec{r(t)}}.
\end{equation}
We use the notation $\vnabla_\veck =(\p/\p k_x,\p /\p k_y)$. 
Solving \eqref{eq:rayeq} numerically  (we use the Dormand-Prince algorithm \cite{DoPri}), 
yields $\veck(\vecx)$ and $\phih(\vecx)$  along a ray defined by $\vecr$. 
The phase field $\phi(\vecx,t)$ is obtained for a rectangular grid in the $\vecx$ plane from
the union of a family of rays $\vecr(t;\alpha)$ starting from locations $\vecr_0(\alpha)=\vecr(0,\alpha)$, such that the 
desired domain is covered.
         
\begin{figure}[ht!]
		\centering
		\includegraphics[width=0.28\textwidth]{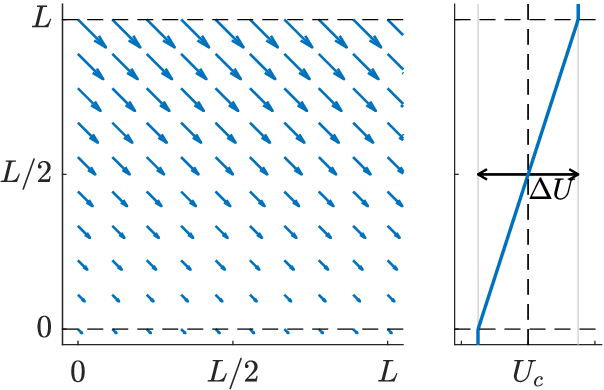}%
        \includegraphics[width=0.19\textwidth]{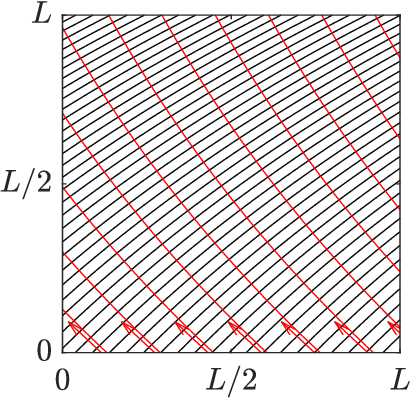}
		\caption{
        Illustration of the change in the wave field due to the presence of horizontal shear. The left and center panel show the  background current $\vecU$ with $\AngU=-\pi/4$ and the right panel shows the rays $\vecr$ (red) for one input wave vector $\veck$ (red arrow) and the crest-lines (black). Note how the rays bend and the wavelength decreases towards $y=L$.
        }
		\label{fig:current and phase illustr}
	\end{figure}

The initial conditions  $\veck_0(\alpha)$ and $\phih_0(\alpha)$ for the wave vector and phase, respectively,  must obey 
\begin{equation}
 \Omega(\vecr_0(\alpha),\veck_0(\alpha))-\omega=0
\end{equation}
and
\begin{equation} \label{eq:IC_phi0}
 \pdiffOne{\phih_0(\alpha)}{\alpha}=\pdiffOne{\vecr_0}{\alpha}\cdot \veck_0(\alpha).
\end{equation}
The effect of horizontal shear on the evolution of the phase and wave vector of a refracted plane wave is illustrated in Figure \ref{fig:current and phase illustr}.

\section{Methods} \label{sec:methods}

\subsection{Dimensional basis}\label{sec:dimensions}
 
Let dimensional quantities be denoted with an asterisk. The basis for non-dimensional quantities is constructed using $g^*$ and a reference wavenumber $\kref^*$ from which we define a length- and a time-scale, $\Lref^*=2\pi/\kref^*$ and $\Tref^*=2\pi/\omref^*$, respectively, where $\omref^*=\sqrt{g^*\kref^*}$. Non-dimensional velocities $U=U^*\Tref^*/\Lref^*=U^*/(\omref^*/\kref^*)$, therefore, are given in units of the quiescent-water phase velocity $\CPref^*=\omref^*/\kref^*$.

\subsection{Normalized scalar product} \label{sec:doppler}
\label{sec:NSP}

Among several options we choose to consider the much-used normalized scalar product (NSP) method for extraction of DSVs from the  $\veck$-$\omega$-spectrum, for the same reasons argued in \cite{Weichert2023}. The principles of the method have a bearing on the biases we will investigate, so we briefly review for them. Further details may be found in, e.g., \cite{Serafino10,Smeltzer19}.

First, the 3D amplitude spectrum is obtained as $F(\veck,\omega) = |\FFT[\zeta(x,y,t)]|$ (FFT: Fast-Fourier transform).
 Then, to fit the dispersion relation with fit parameter $\vecDSV$ to the measured spectrum, we calculate and maximize the scalar product between $F$ and a function $G(\veck,\omega;\vecDSV)$ 
 which is essentially zero except on the dispersion cone $\omega = \omDR(\veck,\vecDSV)$ and a small width around it in the $\omega$ direction (see e.g.\ \cite{Serafino10,Smeltzer21}).
 
It is instructive to use cylindrical coordinates here even though the actual implementation of the method is Cartesian. 
Writing $(\veck,\omega)=(k\cos\Angg,k\sin\Angg,\omega)$, the characteristic function $G$ is
\begin{equation} \label{eq:DSVkernel}
  G(\myvartheta,\omega;\vecDSV) = G^+(\myvartheta,\omega;\vecDSV)+G^-(\myvartheta,\omega;\vecDSV),
\end{equation}
where $G^\pm(\myvartheta,\omega;\vecDSV)= \exp[-\frac2{w^2}(\omega\pm\omega_\text{DR}(\mp\myvartheta;k,\vecDSV))^2 ]$.
(Note that the definition of  $G^\pm$ given in equation (15) in \cite{Weichert2023} contains a sign error for negative frequencies; this was of no consequence there since spectral data at negative frequencies was removed due to redundancy.)
The width parameter $w$ must be chosen carefully;
See discussion in Section \ref{sec:DSVextract}, and Appendix B of \cite{Weichert2023}.

\subsection{Wave spectrum}\label{sec:spectrum}

For the power spectrum of the input waves we assume the form
\begin{equation}\label{eq:specQ}
     \hat{Q}(k,\theta) = Q(k)f(\theta),
\end{equation}
with $k$ being the wavenumber $k=|\veck|$, and $\Angg$ the direction of $\veck$ in the $k_x$-$k_y$-plane, $\veck = k(\cos\Angg,\sin\Angg)^T$. The radial component $Q(k)$ is a top-hat spectrum, constant for $\kmin\leq k\leq \kmax$ and zero outside. 
We find that the shape of $Q$ matters little to our results --- several options were tried for the calculations in Section \ref{sec:results} and produced indistinguishable DSV statistics --- except that a peaked spectrum can lead to biases from spectral leakage separate from those we consider \cite{Weichert2023}, which we wish to eschew. The value of $Q$ is irrelevant for linear waves.

Common shapes for the angular distribution function $f(\Angg)$ include the Mitsuyasu \cite{longuet62,Mitsuyasu1976} $\propto\cos^{2s}(\Angg-\Ang)$ and Donelan \cite{Donelan1985} $\propto\text{sech}^2(\beta(\Angg-\Ang))$.
Another reasonable model shape, used by Socquet-Juglard \emph{et al}.~\cite{SocquetJuglard2005}, is a compact
cosine-square function 
with a width $\DAng$, 
\begin{equation}\label{eq:angularspread}
    f(\Angg)= \left\{ 
    \begin{matrix}			
        \cos^2\left(\pi\frac{\Angg-\Ang}{\DAng}\right), &\, |\Angg-\Ang|\le \DAng/2  \\
        0, &\, \text{otherwise}		
    \end{matrix} \right.. \, \quad \,	
\end{equation}

Wave amplitudes are found from the spectrum via \cite{SocquetJuglard2005}
\begin{equation}\label{eq:amplitudes}
    a(\veck) = \sqrt{2\,k^{-1} \,\hat{Q}(k,\Angg)\delta k}
\end{equation}
on an evenly spaced grid in $k_x$-$k_y$, with spacing $\Delta k_x=\Delta k_y=\Delta k$. Note that $a$ decreases as $k^{-1/2}$ along rays of constant $\Angg$.

\subsection{Wave simulation}\label{sec:simulation}

The synthetic surface elevation data is generated by propagating waves from $y=-\half L$ through the domain defined by $|x|\leq\half L,\,|y|\leq\half L$.  The following steps are followed.
 
(1) For given $\Ang$ and $\DAng$, amplitudes $a(\veck)$ are found from equation \eqref{eq:amplitudes} for each $\veck$ in the grid, thus providing the local wave vector amplitude spectrum at $y=0$, the center of the domain,  as input. Corresponding frequencies are obtained by $\omega(\veck)=\omega_0(k)+\veck\cdot\Uc$ where $\Uc=\vecU(0)$, the central current. We require initial values of $\veck$ and $\phi$ (denoted with subscript `0') at $y=-\half L$; solving equation $\Omega(\vecx,\veck)=\omega$ yields $\veck_0$ for a given $\veck$, and equation \eqref{eq:IC_phi0} gives $\phi_0$ (trivial, since $\veck$ is constant along $y=-\half L$) with the integration constant $\phih(-\half L,-\half L)$ for each wave component picked randomly from a uniform distribution between $0$ and $2\pi$. 
 
 (2)  The phase field $\phih(x,y;\veck)$ is then found for each $\veck$ by solving equation \eqref{eq:rayeq} (see Section \ref{sec:simulation} for details).
 
 (3) The surface elevation is then calculated as
 \begin{equation}
     \zeta(x,y,t) = \sum_{\veck} a(\veck)\cos[\omega(\veck) t -\phih(x,y;\veck)].
 \end{equation}
%

\section{Parameter Choice}\label{sec:parameters}

 Even under the simplifying assumptions of linear horizontal shear and linear wave propagation following the geometric optics approximation, the number of free parameters is large. Here we summarize the parameter choices for the scenarios studied in this work.

\subsection{Domain}

The domain size is a compromise between spectral resolution and computational cost. 
$L=10$ (reference wavelengths), duration $T=40$ (reference wave periods),  resolutions $\delta x = \delta y=1/28$ and $\delta t=1/14$. The domain size $L$ was, in fact found to not have a direct influence on results, provided absolute current changes $\Delta U$ were kept the same, so long as the detrimental effects of finite wavenumber resolutions $\delta k=2\pi/L$ and $\delta\omega=2\pi/T$ were dominated by other sources of uncertainty. 
The implied resolution in DSV extraction from finite $L$ and $T$ are 
$\dcdk=(\del\omega/\del k) \,\delta k/k=\half k^{-3/2}\delta k $ and $\dcdw=\delta \omega /k$, respectively \cite{Smeltzer19}. 

\subsection{Spectrum}
For the input spectrum, equations \eqref{eq:specQ} and \eqref{eq:angularspread}, the mean propagation direction was varied within $\Ang\in[\ang{60} ,\ang{120}]$.  The angular spread was varied within $\DAng\in[\ang{20},\ang{40}]$. For values outside these ranges partial reflection of rays would occur, causing caustics, which are beyond the scope of this work. The values for angular spread represent directionally narrow wave spectra, similar to cases studied in \cite{SocquetJuglard2005} (the precise definition of $\Delta\theta$ varies between works). 
The input  spectrum \eqref{eq:specQ} is defined on an evenly spaced grid with  $\Delta k\approx 0.341/L, \kmin=0.1$, and $\kmax=3.5$.

\subsection{Background current}

We shall consider the simple case where $\vecU$ varies only with $y$ and has constant direction, i.e.,
$\vecU (\vecx) = U(y)(\cos\AngU,\sin\AngU) \equiv U(y) \vece_U$. As in \cite{JoSko78} we let $U$ vary linearly over a range $\Delta U$ through a band of width $L$ at the center of the domain,
\begin{equation}\label{eq:linearcurrent}
   \vecU (y) = \begin{cases}
        \vece_U (U_c-\Delta U/2)\quad   & \text{if } y\le -L/2\\
		\vece_U (U_c+\Delta U\,y/L)\quad   & \text{if } |y|< L/2\\
        \vece_U (U_c+\Delta U/2)\quad   & \text{if } y\ge L/2\\
	\end{cases}.
\end{equation}
The left and center panels in Figure \ref{fig:current and phase illustr} show an example with $\AngU=-\ang{45}$.
We consider the range $0\leq\Delta U\leq0.1$, i.e., $\Delta U$ can reach $10\%$ of the reference-wave phase velocity, and the centerline current (the choice of reference system) is set to $U_c=0.05$. While significant, the current is small enough that the shortest (hence slowest) waves in the spectrum are refracted, not reflected, in all cases considered.
These parameter choices are reasonable for variations in ocean currents; for example, field observations by 
 Romero \emph{et al}.~\cite{Romero2017}, 
Lund \emph{et al}.\ \cite{Lund18}, Vre{\'c}ica \emph{et al}.\ \cite{VreCica2022} and Lenain \emph{et al}.\ \cite{Lenain23} show current variations which, when normalized by the phase velocity of a typical (typical large) wavenumber ($k_\text{mid}^*$ and $k_\text{hi}^*$ in Table \ref{tab:values}), are roughly 
$0.04$ ($0.13$), 
$0.06$ ($0.10$), $0.07$ ($0.17$), and $0.04$ ($0.08$), respectively. It should be noted, however, that 
in flows other than the ocean, e.g.\ estuaries and rivers, changes in current can be far more abrupt, see e.g.\ \cite{Stresser17}. In such cases a more general wave theory is required.

\begin{table}[ht]
    \begin{center}
    \begin{tabular}{lccc}
         \hline
         Reference & $k_\text{mid}^*$ [rad/m] & $k_\text{hi}^*$ [rad/m] & $\Delta U^*$ [m/s] \\
         \hline
         Lund \textit{et al.}~\cite{Lund18}& 0.1 &0.3 & 0.60 \\
         Vre{\'c}ica \textit{et al.}\ \cite{VreCica2022}& 0.06 &0.3 & 0.94\\
         Lenain \textit{et al.}~\cite{Lenain23}& 0.3 &1.5 & 0.20\\
         Romero \textit{et al.}~\cite{Romero2017}
         & 0.08 &1.0 & 0.40\\
         \hline
    \end{tabular}
    \end{center}
    \caption{Characteristic values from field measurements}
    \label{tab:values}
\end{table}
\vspace{-0.5cm}

\subsection{DSV extraction}\label{sec:DSVextract}

To mitigate the effects of spectral leakage
the data $\zeta(x,y,t)$ is tapered with a 3D Hann window (see \cite{Weichert2023}) before FFT is applied to obtain $F(\veck,\omega)$, from which DSVs are extracted.
As will be detailed below, the introduction of horizontal shear causes $F(\veck,\omega)$ to no longer be strongly concentrated near a single dispersion curve, but to be smeared out between an upper and lower dispersion curve corresponding to the lowest and highest current velocities. 
For the DSV extraction algorithm we set the half-width $w$ of the characteristic function $G$ to $w=0.2$. This ensures that $G$ is wide enough to encompass the largest frequency spread $\Delta\omega=0.35$ (due to $k=3.5$ and $\Delta U_\text{max}$) we consider.
Further increasing $w$ had little effect on the DSV results while lower values tended to strengthen biases  and variances compared to those we report below.

\section{Results  and Discussion} \label{sec:results}

In this section we present the results in terms of extracted DSVs from the synthetically generated wave data, for parameter combinations discussed above.

\subsection{The effect of horizontal shear}

    \begin{figure}[ht!]
		\centering
        \includegraphics[width=0.45\textwidth]{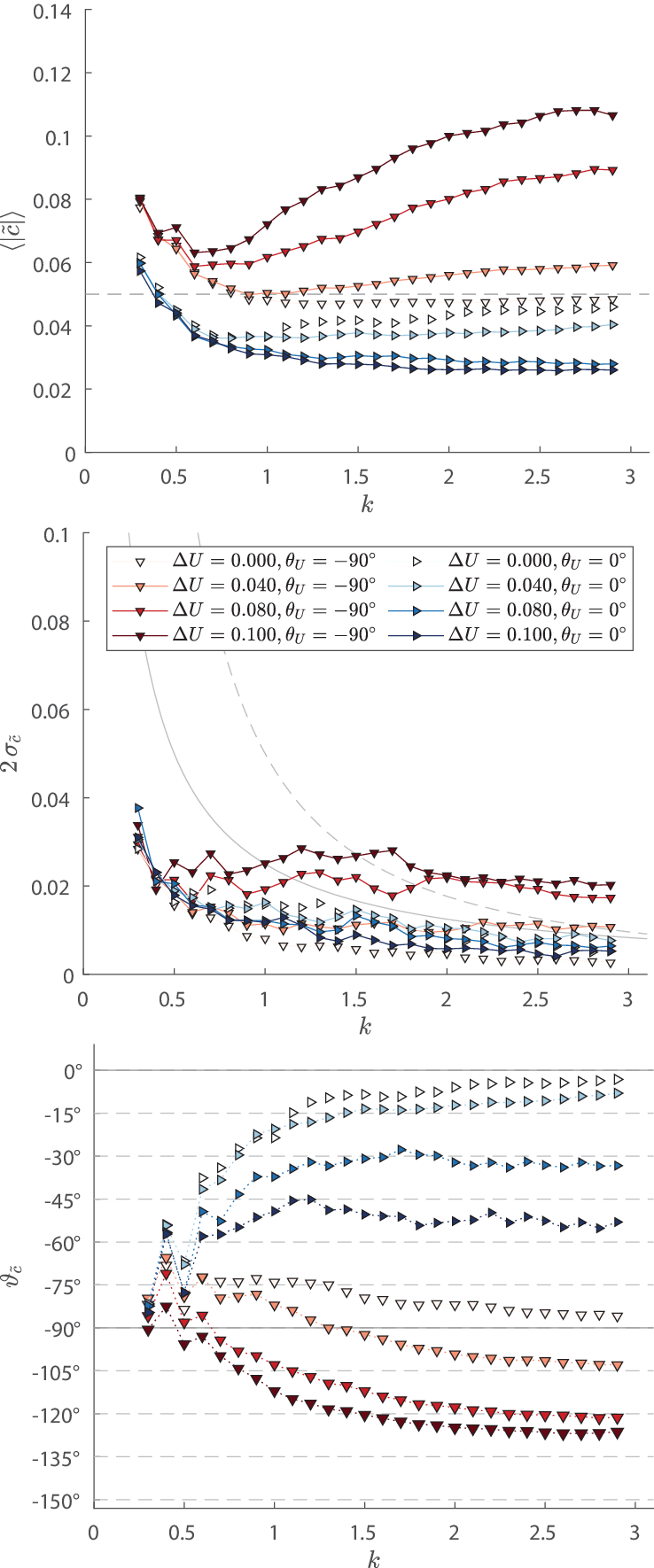}
		\caption{
        Statistics of Extracted Doppler Shift velocities (DSVs) $\vecDSV(k)$ for waves with small angular spread $\DAng=\ang{20}$ atop a linearly sheared current. The current strength varies linearly within $0.05\pm\Delta U/2$ along the $y$-axis, i.e. along $\Angg=\ang{90}$, and has a constant direction of $\AngU=\ang{-90}$ (red downward facing triangles) or $\AngU=\ang{0}$ (blue right facing triangles). The top panel shows the average of the extracted DSVs, the middle panel shows the standard deviation $\sigma_{\DSV}$, times two, and the bottom panel shows the direction of the average DSV. The middle row also shows the implied velocity resolutions $\dcdk$ (dashed) and $\dcdw$ (solid).
		}
		\label{fig:DSVs_example1}
	\end{figure}
    
\subsubsection{An illustrative example} 

In order to illustrate the effects of horizontal shear clearly, we first consider  a scenario with particularly unfavorable parameters: a very narrow angular distribution,  $\DAng=\ang{20}$,  velocity variations up to $\Delta U =0.1$, current direction $\Ang=\ang{120}$, and two different wave propagation directions, $\AngU=\ang{0}$ and $\AngU=\ang{90}$.
Relative errors in these cases reach $100\%$ and $-50\%$, respectively (see Table \ref{tab:directionScan}).

The wavenumber-resolved DSVs $\vecDSV(k)$ for these cases are shown in Figure \ref{fig:DSVs_example1} in terms of the statistical average $\myMean{|\vecDSV|}$ (top), its standard deviation $\sigma_{\DSV}$ (middle) and the average direction $\AngDSV$ of the extracted DSV vector (bottom). Values other than $\myMean{|\vecDSV|}=0.05$ and $\AngDSV=\ang{0}$ or $\ang{90}$ constitute a bias.
    The influence of horizontal shear is obvious  as the biases increase both with $k$ and with $\Delta U$,  over- or underestimating the domain-averaged current by up to a factor of $2$.
    This worsening is contrary to what would otherwise be expected, since the implied resolutions $\dcdk$ and $\dcdw$ improve with higher $k$.

    \begin{figure}[ht]
        \begin{center}
            \includegraphics[width=.49\textwidth]{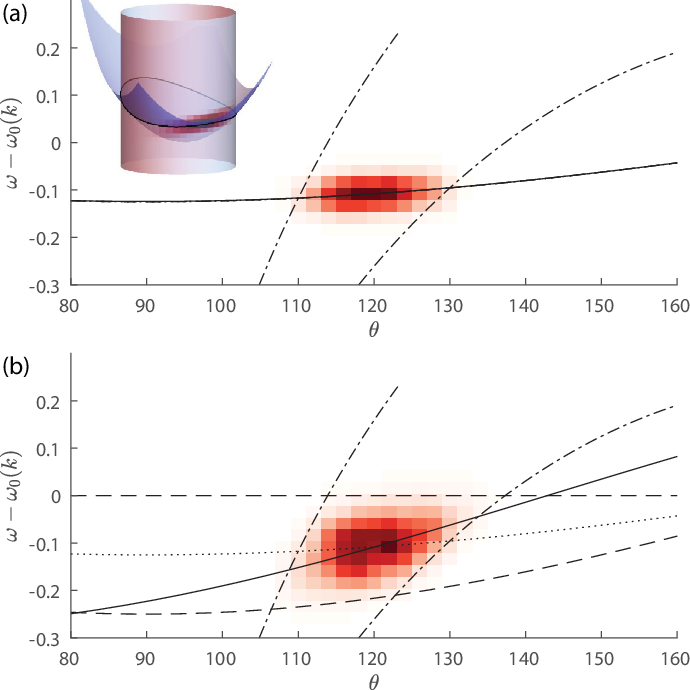}
        \end{center}
        \caption{Visualization of the origin of biases originating from horizontal shear.  The panels show the spectral intensity $F(k\cos\Angg,k\sin\Angg,\omega)$ on the cylinder surface with radius $k=2.5$ for an input spectrum of width $\DAng=\ang{20}$ and mean propagation direction $\Ang=\ang{120}$. The solid line shows the best fit for the dispersion relation on this cylinder surface. (a) Uniform current; (b) horizontally sheared current with same average current as in (a). The dotted line is the dispersion relation for the uniform/average current. In both panels the dash-dotted  and dashed lines show the limiting curves obtained with equations \eqref{eq:proj_spec_line} and \eqref{eq:omlimnew}, respectively. 
		}
		\label{fig:unrolCylExamples}
	\end{figure}

\subsubsection{The origin of biases}     
    Explaining how $\Delta U>0$ can lead to these biases requires a closer look at the shape of the spectrum that enters the DSV extraction step detailed in Section \ref{sec:NSP}. In very broad strokes, the following occurs. 
     \begin{itemize}
         \item For each value of $|\veck|$ the NSP procedure considers the spectral signal on a cylindrical surface $|\veck|=\kcyl$ in $\veck$-$\omega$ space, and, assuming the signal is concentrated near the dispersion relation for an unknown uniform current, fits it to the dispersion relation $\omega(\theta)=\omDR(\veckcyl(\theta),\vecU)$ with $\vecU$ as fitting parameter (see Figure 4 of \cite{Weichert2023}). [Here: $\veckcyl(\theta)=(\kcyl\cos\theta,\kcyl\sin\theta)$]
         \item If the current is not in fact uniform, the spectral signal remains centered near the dispersion relation corresponding to a uniform current equal to the mean current, $\omega(\theta)=\omDR(\veckcyl(\theta),\vecU_c)$, but is more spread out on the cylinder surface; we refer to this as smearing. The smeared signal is in general asymmetric about the mean-current dispersion curve. 
         \item For highly directional wave spectra at some oblique angles, the asymmetric smearing causes the best fit for $\omega(\theta)$ to deviate significantly from that corresponding to the average current, and a DSV systematically different from $\vecU_c$ is extracted.
     \end{itemize}
     The process is illustrated in Figure \ref{fig:unrolCylExamples}, and details now follow.

    Bear in mind that the input spectrum \eqref{eq:specQ} has the separated form $\hat{Q}=Q(k) f(\theta)$ and is symmetrical in $\theta$ around the main direction $\theta_0$ and fully contained within a sector of width $\Delta\theta$. With smearing, the measured spectrum $F$ does not have these properties. The origin of biases due to shear lies in the sequence of steps that transforms the initially symmetric input spectrum into the FFT signal which the algorithm uses to estimate DSVs.

     Consider a single wave whose wave vector at the center of the domain is $\veckap$.  In the case of a uniform current $\vecU=\Uc$, its contribution to the spectrum $F$ is concentrated near the single point $(\veck,\omega) = (\veckap,\omDR(\veckap;\Uc))$.   When the current varies across the domain as a function of $y$, however,   the wave vector component $k_y$ varies with $y$, while $\omega$ and $k_x$ are the same throughout the domain. The spectral contribution from $\veckap$ is instead spread along a line segment spanning a finite range of $k_y$ values. 

     We now study how smearing changes the appearance of the spectral signal which the NSP algorithm uses for curve fitting. Consider the spectrum $F(\veckcyl(\theta),\omega)$ for a cylinder surface $k=\kcyl$. 
     Theoretically, without horizontal shear, only wave components with wavenumber $k=\kcyl$ lead to spectral intensity on this surface. However, since $k_y(y)$ now varies over the domain, but $k_x$ does not, the FFT-signal on the cylinder now comes from waves with a range of wavenumbers different from $k$, and hence is smeared over a range of frequencies
     \begin{equation}
         \omega_{\min}
         \leq \omDR(\veck,\Uc) \leq \omega_{\max}.
     \end{equation}

    To determine the upper and lower frequency-smearing bounds, assume $\veckCP = (\kCPx,\kCPy)$ is the wave vector obtained by projecting a wave vector $\veck$ in the shear direction (here: $y$) onto the cylinder surface; for $|k_x|\leq \kcyl$, we then have $\kCPx=k_x$ and $\kCPy = (\kcyl^2-k_x^2)^{1/2}\mathrm{sign}(k_y)$.
    For the signal from a wave  $\veck$ to reach the cylinder surface, the range of frequencies of $\veckCP(y)$ must contain the frequency $\omega(\veck)$, hence, 
     \begin{align} \label{eq:omlim1}
        \begin{split}
            \omega_\text{min} &= \min_y\{\omDR(\veckCP,\vecU)\} ;  \\ 
            \omega_\text{max} &= \max_y\{\omDR(\veckCP,\vecU)\}   ,
        \end{split}
    \end{align}
    and $\min_y$ ($\max_y$) are the minimum (maximum) of $\omDR(\veckCP,\vecU(y))$ varying $y$ throughout the domain while keeping $\veckCP$ constant.

    Such a broadening of $F$ intuitively leads to to higher curve-fitting uncertainty, but not necessarily systematic bias. To see the latter one should explain how the smeared spectral signal also becomes skewed. Specifically, the NSP method uses the spectrum in the $\omega$-$\theta$ plane resulting from unrolling the cylinder $k=\kcyl$. 
            
    Consider the spectral signal on a cylinder $k=\kcyl$ from wavevectors along a line of constant angle $\ttheta$ in the $\veck$-plane of the input spectrum (defined at $y=0$), parameterized by wavenumber $\onewave$, i.e., $\veck=\veckap=(\onewave\cos\ttheta,\onewave\sin\ttheta)$.
    These wave components have a (constant) frequency $\omDR(\veckap,\Uc)$. Since $U(y)$ varies, spectral signal is left on the cylinder from a range of $\onewave$ either side of $\kcyl$, drawing a curve In the $\theta$-$\omega$ plane resulting from unrolling the cylinder surface, a curve is drawn given by
    \begin{equation} \label{eq:proj_spec_line}
            \vectwo{\Angg}{\omega}(\onewave)
            =\vectwo{\arccos{\left(\frac{\onewave}{\kcyl}\cos\ttheta\right)}}{\omDR(\veckap,\Uc)}.
    \end{equation}
    In words, a signal observed at angle $\theta$ is in general caused by waves propagating in directions $\ttheta\neq\theta$.
            
    The spectral signal is circumscribed by the curves \eqref{eq:proj_spec_line} due to  $\ttheta=\theta_0\pm\half \Delta\theta$ (dash-dotted lines) and the frequency limits of equation \eqref{eq:omlim1}. Since the direction of the current is constant, $\omega_\text{max}$ and $\omega_\text{min}$ are, respectively, the greater and smaller value of $\omega_+(\theta)$ and $\omega_-(\theta)$, given by
    \begin{equation}\label{eq:omlimnew}
        \omega_\pm(\theta) =  \omDR(\veckCP,(U_c\pm\half\Delta U)\vece_U)
    \end{equation}
    with $\vece_U=\vecU/U$. These limiting frequencies are shown as dashed curves in Figure \ref{fig:unrolCylExamples}. Since all the bounding curves are projections of constant-$\ttheta$ into the $\theta$-$\omega$ plane, they are all curved, and the smeared spectrum is skewed.
            
    The geometry described can be seen in Figure \ref{fig:unrolCylExamples}, wherein a comparison is shown between the spectrum on a cylinder surface of radius $\kcyl=2.5$ with and without shear for our previously considered spectrum with $\DAng=\ang{20}$ and $\Ang=\ang{120}$. 
    We can note two effects: The blurring of the spectrum in the $\omega$ direction increases the susceptibility to spectral noise, leading to  exacerbated random biases. (The spectra in Figure \ref{fig:unrolCylExamples} are averaged spectra, not showing noise). The second effect is that, because of the skewness of the smeared signal, the best fit of a cosine to the FFT signal, shown as a solid line, obtains a different slope than the ``correct'' curve (dotted line). This leads to biases in both DSV magnitude and direction, strongly dependent on the position of the spectral intensity on the cylinder and the current direction.
    Clearly the severity of the biases will generally be worse the smaller $\Delta\theta$ is: the broader the  angular spread the closer the best fit will be to the average-current dispersion curve.

    \subsection{Direction dependence}\label{sec:directiondependence}
     From Figure \ref{fig:unrolCylExamples} and  equations \eqref{eq:proj_spec_line} and \eqref{eq:omlimnew}  it is clear that the shape of the area of spectral intensity on an unrolled cylinder depends on the direction of the current $\AngU$, and the range of directions of the spectral components $\Ang \pm\DAng/2$.
      Therefore, extracted DSVs for no-shear and the maximum shear $\Delta U=0.1$ are extracted and compared for a variety of propagation directions $\Ang$, angular spreads $\DAng$ and current directions $\AngU$. The background current $\vecU = U(y)\vece_U$ follows equation \eqref{eq:linearcurrent} as before.
      We let $\nabla U = U'(y) \vece_y$ and denote $U'(y)\geq 0$ the `magnitude gradient'.
      
      The results at $k=2.5$ are shown in Table \ref{tab:directionScan}. While the precise numbers in the table depend on the precise geometry and exact method of DSV extraction, and have some variance due to the finite number of simulations, they indicate trends and order of magnitude of biases.
        
        \begin{table}[ht]
            \centering
\newcommand{\stackvals}[3]{\hspace{-0.2cm}\begin{tabular}{c}#1\\#2\\#3\\\end{tabular}\hspace{-0.2cm}}
\begin{tabular}{|c|c|c|c|c|c|}
\hline
\diagbox{$\Ang$}{$\AngU$}
&\ang{                 -90}  &\ang{                 -45}  &\ang{                   0}  &\ang{                  45}  &\ang{                  90}\\       \hline
\ang{                  60},\, \stackvals{$\DAng=\ang{20}$}{$\DAng=\ang{30}$}{$\DAng=\ang{40}$}  &\stackvals{111.4}{51.0}{24.7}  &\stackvals{3.2}{0.1}{-1.3}  &\stackvals{48.8}{29.6}{22.2}  &\stackvals{105.3}{46.1}{26.5}  &\stackvals{20.0}{-9.7}{-10.5}\\  \hline
\ang{                  90},\, \stackvals{$\DAng=\ang{20}$}{$\DAng=\ang{30}$}{$\DAng=\ang{40}$}  &\stackvals{0.3}{-3.0}{-3.1}  &\stackvals{-14.5}{-9.5}{-8.0}  &\stackvals{-17.0}{-7.9}{-5.1}  &\stackvals{-3.7}{1.6}{3.2}  & \stackvals{6.5}{4.0}{3.2}\\  \hline
\ang{                 120},\, \stackvals{$\DAng=\ang{20}$}{$\DAng=\ang{30}$}{$\DAng=\ang{40}$}  &\stackvals{\textbf{108.2}}{49.6}{26.0}  &\stackvals{81.6}{22.1}{0.8}  &\stackvals{\textbf{-48.5}}{-36.5}{-28.1}  &\stackvals{-27.3}{-14.3}{-7.6}  &\stackvals{20.0}{-9.3}{-10.6}\\  \hline
\end{tabular}
            \vspace{1mm}
            \caption{Maximum deviation in velocity magnitude of extracted DSVs from the domain average $U_c$, in percent of $U_c$. Here $k=2.5$ and the current is equation \eqref{eq:linearcurrent} with $\Delta U=0.1$, hence $U_c=0.05$.
            The cases shown in bold are particularly unfavourable combinations which we discuss in more detail; see Section \ref{sec:shearDependence} and Figure \ref{fig:DSVs_DU_scan}.
            }
            \label{tab:directionScan}
        \end{table}

         Unless the waves propagate along the shear direction, i.e. $\Ang=\ang{90}$, the biases when $\Delta \theta=\ang{20}$ are strong enough to make the extracted DSVs practically unusable. This is strongly mitigated by increasing the angular width $\DAng$ of the wave spectrum, in some cases nearly eliminating the biases, although deviations from the mean in excess of $25\%$ remain also for $\Delta\theta=\ang{40}$ in some cases. The dependence on $\theta_0$ and $\theta_U$ is very strong. 
         The largest biases are overestimations, more than $100\%$ for extremely narrow spectra. 
         Large underestimations occur less often, but are nearly equally severe, reaching $-50\%$.

    \subsection{Shear dependence}\label{sec:shearDependence}
    
    \begin{figure}[ht!]
        \centering
        \includegraphics[width=0.45\textwidth]{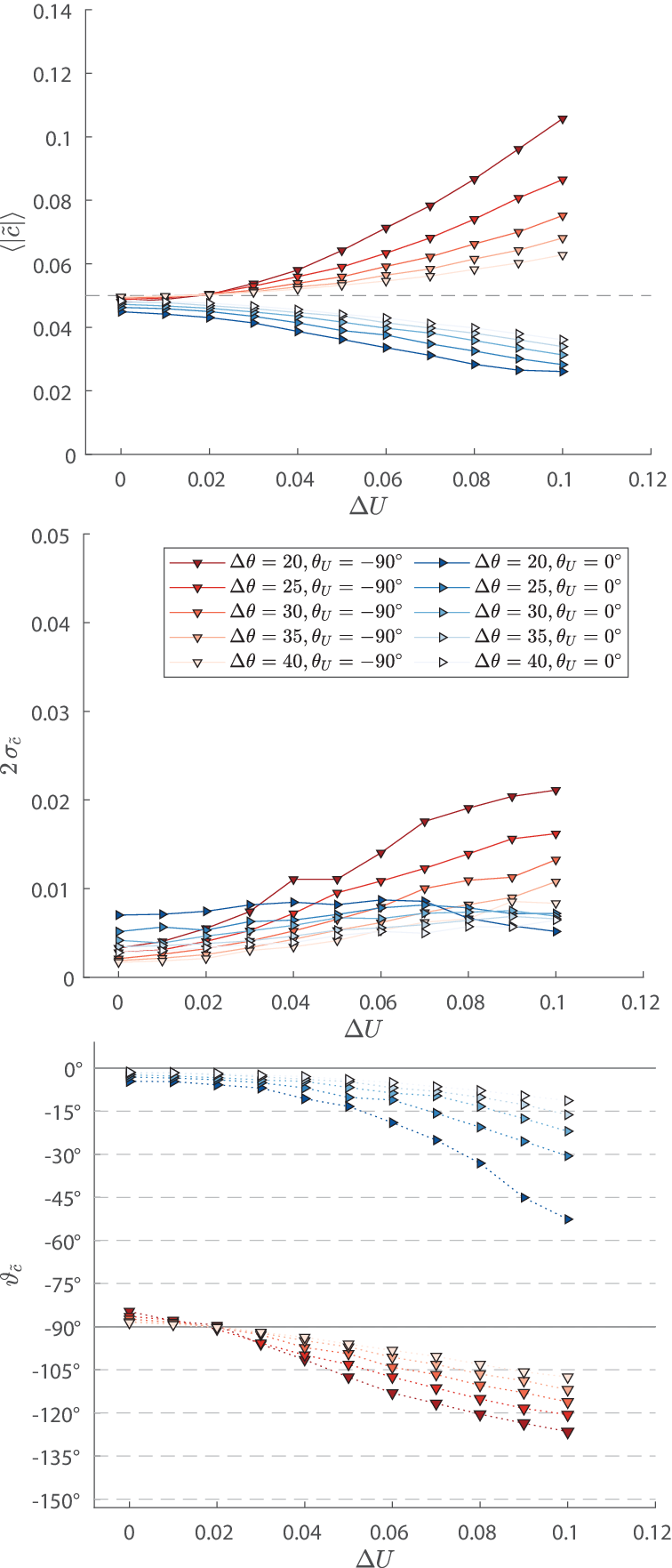}
        \caption{ 
            Influence of horizontal shear and angular spread on the extraction of Doppler Shifts. Waves with different angular spreads $\DAng$ propagate atop a linearly sheared current. The current strength varies linearly within $0.05\pm\Delta U/2$ along the $y$-axis, i.e. along $\Angg=\ang{90}$, and has a constant direction of $\AngU=\ang{-90}$ (red downward facing triangles) or $\AngU=\ang{0}$ (blue right facing triangles). The data presented here is obtained from results like those shown in Figure \ref{fig:DSVs_example1}, but for one wavenumber, $k=2.5$.  As in Figure \ref{fig:DSVs_example1} the top panel shows the average extracted DSV, the middle row the standard deviation and the bottom panel the direction of the average DSV. 
        }
        \label{fig:DSVs_DU_scan}
    \end{figure}
    
    In section \ref{sec:directiondependence} we showed that the presence of shear can lead to both over- and underestimating the DSVs and that a larger angular spread in the spectrum mitigates these. For two scenarios with particularly large biases ([$\Ang, \AngU$]=[$\ang{120},\ang{90}$] and [$\ang{120}, \ang{0}$]) we 
    vary $\Delta U$ between $0$ and $0.1$  and $\DAng$ between $\ang{20}$ and $\ang{40}$.
    The statistical average, standard deviation and direction of the extracted DSVs at $k=2.5$ are shown in Figure \ref{fig:DSVs_DU_scan} (top, middle and bottom panels, respectively).

    The most noteworthy feature is perhaps that the growth of mean biases $\langle|\DSV|\rangle-U_c$ and $\vartheta_{\DSV}-\theta_U$ increase faster than linearly with increasing $\Delta U$  for moderate ($\Delta U\le 0.05$) shear. This offers the potential for testing whether DSVs exhibit biases from linear shear, because a reduction in domain size will disproportionately reduce the biases.

    \subsection{``Ramp'' current profile within the domain}

    Figure \ref{fig:unrolCylExamples}(b) shows how a changing current causes the spectral signal to be distributed between the limiting curves $\omega_{\max/\min}(\Angg_c)$ which correspond to the highest and lowest velocities in the domain. Generally, the more focused the spectrum is around the dotted curve which represents the mean current, the better with respect to bias and variance, and vice versa. Because the wave data was tapered before DSV extraction, wave amplitudes near the edges of the domain were much reduced, and since this is where the extreme velocities are found, the spectral signal is very weak near the dashed lines in figure \ref{fig:unrolCylExamples}(b).

    The opposite effect occurs if the current changes only in the central part of the domain (say). Since velocities close to the maximum and minimum values now cover a greater part of the domain area, the spectral signal is pushed away from the dotted mean-current curve in figure \ref{fig:unrolCylExamples}(b) and towards the dashed curves, thereby increasing the risk of biases.

\begin{figure}[ht!]
		\centering
        \includegraphics[width=0.45\textwidth]{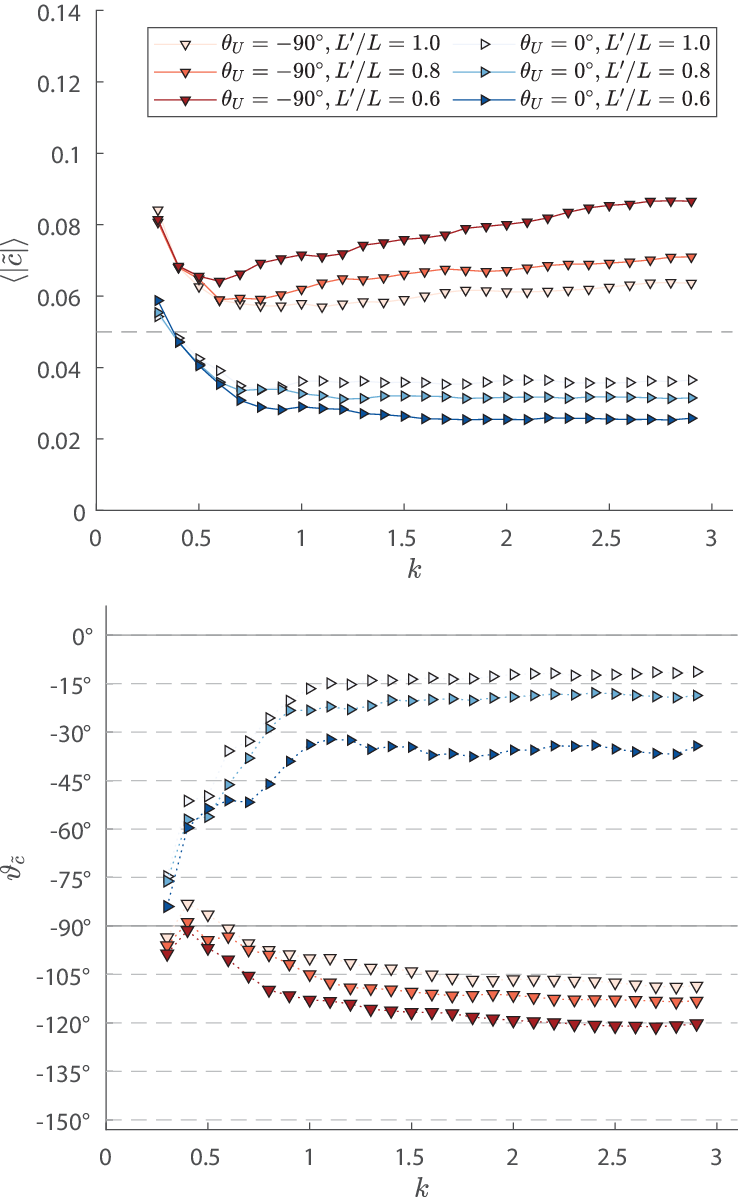}
		\caption{Same as the top and bottom panel of Figure \ref{fig:DSVs_example1}, respectively, with $\Ang=\ang{120}$, $\DAng=\ang{40}$, $\Delta U=0.1$,  and two different $\theta_U$, for three different ramp widths $L'$.
		}
		\label{fig:DSVs_plateau}
	\end{figure}

    We repeated our analysis using a ``ramp'' velocity profile, i.e., replacing $L$ in equation \eqref{eq:linearcurrent} by a ramp width $L'\leq L$. Figure \ref{fig:DSVs_plateau} shows that biases in both magnitude and direction worsen when $L'/L$ decreases, as can be expected from the above, while other trends remain the same.

\section{Conclusions}\label{sec:conclusions}

We have investigated biases in remote sensing of near surface currents due to horizontal current shear. 
A bias occurs when the current extracted using the wave field in a domain differs from the domain average, $U_c$.
Synthetic random wave elevation data were generated in a square domain $-\half L\leq x,y\leq \half L$ from a prescribed, directional wave spectrum and propagated 
by means of raytracing, on a horizontal current $\vecU=U(y) \vece_U$ varying around a mean value $U_c$ in the range $-\half\Delta U\leq U(y)-U_c\leq \half \Delta U$. 

Doppler shift velocities (DSVs, symbol $\vecDSV$) were then extracted from the wave field by means of the much-used normalized scalar product method, while varying the following scenario defining parameters:
\begin{itemize}
    \item Current velocity change $\Delta U$;
    \item Current direction, $\vece_U$;
    \item Mean wave propagation direction;
    \item Angular spread of the directional wave spectrum.
\end{itemize}
The current magnitude gradient is $U'(y)\vece_y$ and we keep $U_c$ constant.

Our investigation yields the following main observations:
\begin{itemize}
    \item Horizontal shear can lead to severe biases in wave-derived currents when the wave spectrum is highly directional, i.e., it has small angular spread; for broader angular spectra the effects of horizontal shear are weaker.
    \item Biases in the magnitude and direction of $\vecDSV$ can both increase faster-than-linearly with increasing $\Delta U$.
    \item Biases are determined by the absolute amount by which the current velocity changes within the field of view, $\Delta U$, more than its local rate of change.
    \item Biases are highly sensitive to the relative direction between waves, currents and the current's direction of change (between $+100\%$ and $-50\%$ in our cases). No clear trend was discovered, and biases are hard to predict \emph{a priori}.  
\end{itemize}
When the current changes only in the central part of the domain, ``ramped up" in the range $|y|< \half L'$ rather than the full domain size, biases tend to increase with other trends remaining the same.

\subsection{Recommendations for detection and mitigation}

Differences in velocities from point to point are often seen in current measurements using DSVs from surface waves, e.g., \cite{Lenain23}, signifying the presence of horizontal shear.
 To detect whether the shear could be biasing the extracted DSVs in such an area, we suggest the following two steps be taken if possible. First, the angular spread of the spectrum is estimated to determine whether the directional distribution is sufficiently narrow, e.g.\ $\leq \ang{40}$ compared to a $\cos^2$ distribution, equation \eqref{eq:angularspread}. If so, the spatial wave domain could be  divided into four sub-domains, and the analysis repeated for a range of  high wavenumbers that are little affected by the reduction in $k$-resolution due to the smaller domain size.  If the DSVs from the full domain differ significantly from the average of the four quarter domains, shear-induced biases should be suspected, and a closer look at the spectrum, similar to Figure \ref{fig:unrolCylExamples} is warranted.

\section*{Acknowledgments}

\noindent The authors would like to thank Dr.\ Luc Lenain for discussions.

\section*{Data availability}
The data that support the findings of this study are available from the authors upon request. All necessary information for generating the underlying data is provided in the paper. 

\bibliographystyle{IEEEtran}
\bibliography{references.bib}

\end{document}